\documentclass[prl,reprint,doublecolumn,superscriptaddress, nobalancelastpage]{revtex4-1}

\usepackage{tikz}

\usepackage{etoolbox}
\patchcmd{\section}
  {\centering}
  {\raggedright}
  {}
  {}
\patchcmd{\subsection}
  {\centering}
  {\raggedright}
  {}
  {}

\usepackage{tcolorbox}
\usepackage[sort&compress]{natbib}
\usepackage{soul}
\usepackage{graphicx}
\usepackage{epstopdf}
\usepackage{newtxmath}

\usepackage{verbatim}
\usepackage{float}
\usepackage{sidecap}
\sidecaptionvpos{figure}{c}
\usepackage{lipsum}
\usepackage{setspace}
\usepackage{dcolumn}
\usepackage{bm}
\usepackage{amssymb}
\usepackage{amsmath}	
\usepackage{mathtools}
\usepackage{color}
\usepackage{multirow}
\usepackage{lipsum}
\usepackage[a4paper]{geometry}
\newgeometry{top=1.25cm,right=1cm,left=1.5cm,bottom=1.25cm}
\usepackage[colorlinks=true, urlcolor=blue, pdfborder={0 0 0}]{hyperref}
\hypersetup{
linkcolor=blue,
citecolor=blue
}

\usepackage{mathtools}

\DeclarePairedDelimiter\ket{\lvert}{\rangle}
\DeclarePairedDelimiterX\braket[2]{\langle}{\rangle}{#1 \delimsize\vert #2}

\usepackage{graphicx}
\usepackage{bm}
\usepackage{hyperref}

\begin{document}

\title{An exchange-based surface-code quantum computer architecture in silicon}

\author{Charles D. Hill}
\email{cdhill@unimelb.edu.au}
\affiliation{School of Physics, The University of Melbourne, Parkville, 3010, Australia}
\affiliation{School of Mathematics and Statistics, The University of Melbourne, Parkville, 3010, Australia}

\author{Muhammad Usman}
\email{musman@unimelb.edu.au}
\affiliation{Centre for Quantum Computation and Communication, School of Physics, The University of Melbourne, Parkville, 3010, Australia}
\affiliation{School of Computing and Information Systems, Melbourne School of Engineering, The University of Melbourne, Parkville, 3010, Australia}

\author{Lloyd C.L. Hollenberg}
\email{lloydch@unimelb.edu.au}
\affiliation{Centre for Quantum Computation and Communication, School of Physics, The University of Melbourne, Parkville, 3010, Australia}

\begin{abstract}
Phosphorus donor spins in silicon offer a number of promising characteristics for the implementation of robust qubits. Amongst various concepts for scale-up, the shared-control concept takes advantage of 3D scanning tunnelling microscope (STM) fabrication techniques to minimise the number of control lines, allowing the donors to be placed at the pitch limit of $\geq$30 nm, enabling dipole interactions. A fundamental challenge is to exploit the faster exchange interaction, however, the donor spacings required are typically 15 nm or less, and the exchange interaction is notoriously sensitive to lattice site variations in donor placement. This work presents a proposal for a fast exchange-based surface-code quantum computer architecture which explicitly addresses both donor placement imprecision commensurate with the atomic-precision fabrication techniques and the stringent qubit pitch requirements. The effective pitch is extended by incorporation of an intermediate donor acting as an exchange-interaction switch. We consider both global control schemes and a scheduled series of operations by designing GRAPE pulses for individual CNOTs based on coupling scenarios predicted by atomistic tight-binding simulations. The architecture is compatible with the existing fabrication capabilities and may serve as a blueprint for the experimental implementation of a full-scale fault-tolerant quantum computer based on donor impurities in silicon.
\end{abstract}

\pacs{Valid PACS appear here}
\maketitle

\section{Introduction}
\noindent
Quantum computing based on spin qubits formed by phosphorus donors in silicon~\cite{Kane_Nature_1998} is an attractive approach for large scale implementation of quantum information processing. Some of the milestones achieved to date include single shot spin readout~\cite{Morello_Nature_2010}, the demonstration of single qubits based on both electron~\cite{Pla_Nature_2012} and nuclear~\cite{Pla_Nature_2013} spins, the fabrication of donor based devices in silicon based on scanning tunnelling microscope (STM) techniques~\cite{Fuechsle_NN_2012, Weber_Science_2012}, the post-fabrication pinpointing of their locations in silicon with the exact lattice site precision~\cite{Usman_NN_2016}, and a direct two-electron SWAP operation~\cite{He_Nature_2019}. With ongoing experimental efforts focused on increasing the number of qubits in quantum devices and achieving control with high fidelities, the challenges associated with scale-up and the design of a universal quantum computer architecture incorporating quantum error correction come into sharper focus.

The development of topological quantum error correction (TQEC) codes such as the surface code has provided a scheme for error correction with a relatively high threshold that is commensurate with experiments~\cite{Bravyi_arXiv_1998, Kitaev_JMP_2002, Raussendorf_NJP_2007, Wang_PRA_2011}. While the physical requirements of the surface code are relatively straightforward to contemplate a two dimensional array of nearest-neighbour coupled qubits. However, for all physical qubit platforms, even with assumptions about quantum interconnects~\cite{Nguyen_SP_2017}, the challenges inherent in the spatial arrangement of gates, and temporal characterisation and control complexity for large numbers of  independent qubits to carry out TQEC are formidable. Since Kane's original concept for a 1D qubit array~\cite{Kane_Nature_1998}, a number of proposals have been presented addressing scalability issues, particularly with respect to the requirements of incorporating quantum error correction~\cite{Hill_science_2015, Pica_PRB_2016, Gorman_npjQI_2016, Tosi_NatureComm_2017, Cai_Quantum_2019}. In Ref.~\cite{Hill_science_2015}, a surface-code architecture was reported for impurity spins in silicon which was based on the dipole interactions between the P impurities. This work presented a detailed design introducing shared control, however it was limited to dipole couplings which are of the order of kHz. The difficulty of providing fast, available couplings in solid state architectures has led to several proposals. Pica \textit{et al}.~\cite{Pica_PRB_2016} proposed a surface code architecture, in which electrons were shuttled between neighbouring qubits. Gorman \textit{et al} addressed the problem of coupling by mechanically moving a set of probe qubits in order to establish the required couplings \cite{Gorman_npjQI_2016}. Tosi \textit{et al} \cite{Tosi_NatureComm_2017} proposed the use of long range couplings provided by a flip-flop qubit, a combination of electronic and nuclear spin states that can be controlled with a microwave electric fields. For donor spins in silicon, the incorporation of exchange interaction in surface-code based error correction schemes is still an open question.

\begin{figure*}[htbp]
\begin{center}
\includegraphics[width=19cm]{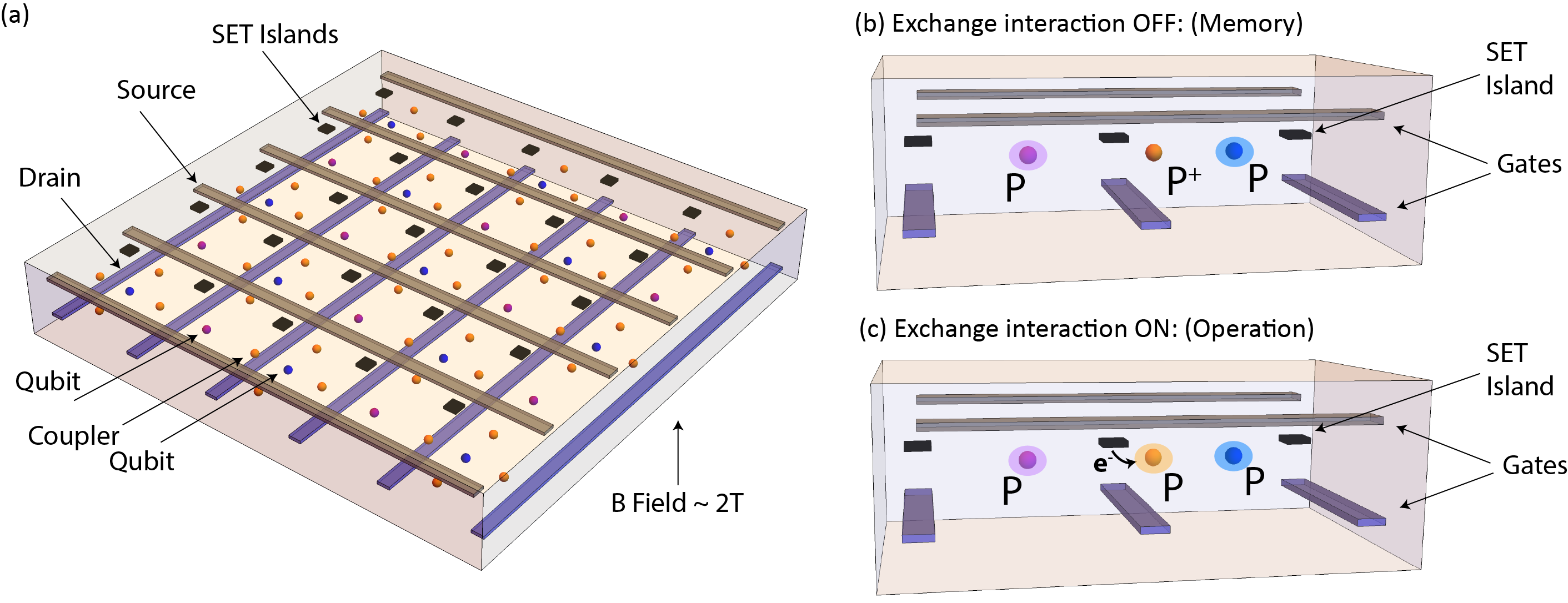}
\caption{\textbf{3D Surface-code Architecture:} The schematic diagram plots the layout of the proposed surface-code architecture based on phosphorus (P) donor qubits in silicon. The architecture is based on previously published scheme~\cite{Hill_science_2015}, however it is updated to exploit the fast exchange interaction between P donor electron spins. The data qubits are separated by 32 nm and additional coupler qubits (orange dots) are incorporated in-between data qubits to control (turn on/off) interaction between them. The qubit plane is addressed by top and bottom gates shown by the blue and gray stripes.} 
\label{fig:architecture}
\end{center}
\end{figure*}

The introduction of shared control~\cite{Hill_science_2015} in donor qubit architecture design space reduces the spatial complexity and dovetails naturally with the repetitive spatio-temporal control requirements of surface code TQEC. Assuming a high level of qubit uniformity and a fundamental qubit pitch of $\geq$ 30 nm, corresponding to the fundamental contol line pitch limit in these devices,  CNOT gates were based on the donor electron spin dipole interaction with a phase-matched electron loading protocol to rectify timing variations associated with the hyperfine interaction. Ideally, one would use the exchange interaction, however, the severe spacing requirements ($\leq$ 15nm) and variations in the exchange coupling work against the design of a direct exchange based 2D array for TQEC. Here, we address these problems by introducing an intermediate donor acting as an exchange coupler. The qubit donors containing quantum data can be spaced comfortably with respect to a control line pitch of 35 nm, and phase matched loading at qubit donors is no longer required. Atomic level simulations, with typical placement variations expected in STM fabrication, indicate CNOT gate times at O($\upmu$sec) are possible and the overall scheme has potential to meet the stringent control requirements of the surface code.

\section{Results \& Discussions}

\subsection{Overview of the Architecture}
\noindent
Figure~\ref{fig:architecture} schematically illustrates the layout of the exchange-based surface-code architecture proposed in this work. The architecture, as its predecessor dipole-based architecture~\cite{Hill_science_2015}, is based on three-dimensional layout. In Figure~\ref{fig:architecture} (a) The colored dots indicate 2D arrangement of donor atoms, interleaved with black squares representing SET islands for loading/unloading and readout of electron to/from qubits. The nuclear spins on donors define the qubit states as shown in Figure~\ref{fig:architecture} (b). The 2D qubit plane is sandwiched between the top and bottom layout of wires forming source and drain. The exponential decay of the exchange interaction with the separation between the donor atoms is well known in the literature, as is the sensitivity of the interaction to valley interference effects \cite{Cullis_PRB_1970, Wellard_PRB_2003, Gonzalez_Nanoletters_2014, Hu_PRB_2005, Sarma_SSC_2005, Wellard_JPCM_2004, Kettle_PRB_2006, Kettle_JPCM_2004, Koiller_PRB_2004, Song_APL_2016, Wellard_Hollenberg_PRB_2005, Testolin_PRA_2007, Saraiva_arx_2014, Pica2_PRB_2014, Koiller_Adabdc_2005, Voisin_2020, Usman_CMS_2021}. This results in a tension between donor separation and exchange strength to design a fast CNOT gate while maintaining sufficient distance between the atoms to allow for control wires, also known as pitch problem. In the previous dipole-based architecture~\cite{Hill_science_2015}, the separation between the adjacent donor atoms was taken to be 30 nm, defined by the gate-leakage pitch limit for STM control-lines. At such distances, the exchange interaction is effectively zero. In our scheme we introduce a coupler donor which switches the exchange on and off by loading and unloading an electron to that position (Figure~\ref{fig:architecture} (c)).   

\begin{figure*}[htbp]
\begin{center}
\includegraphics[width=14cm]{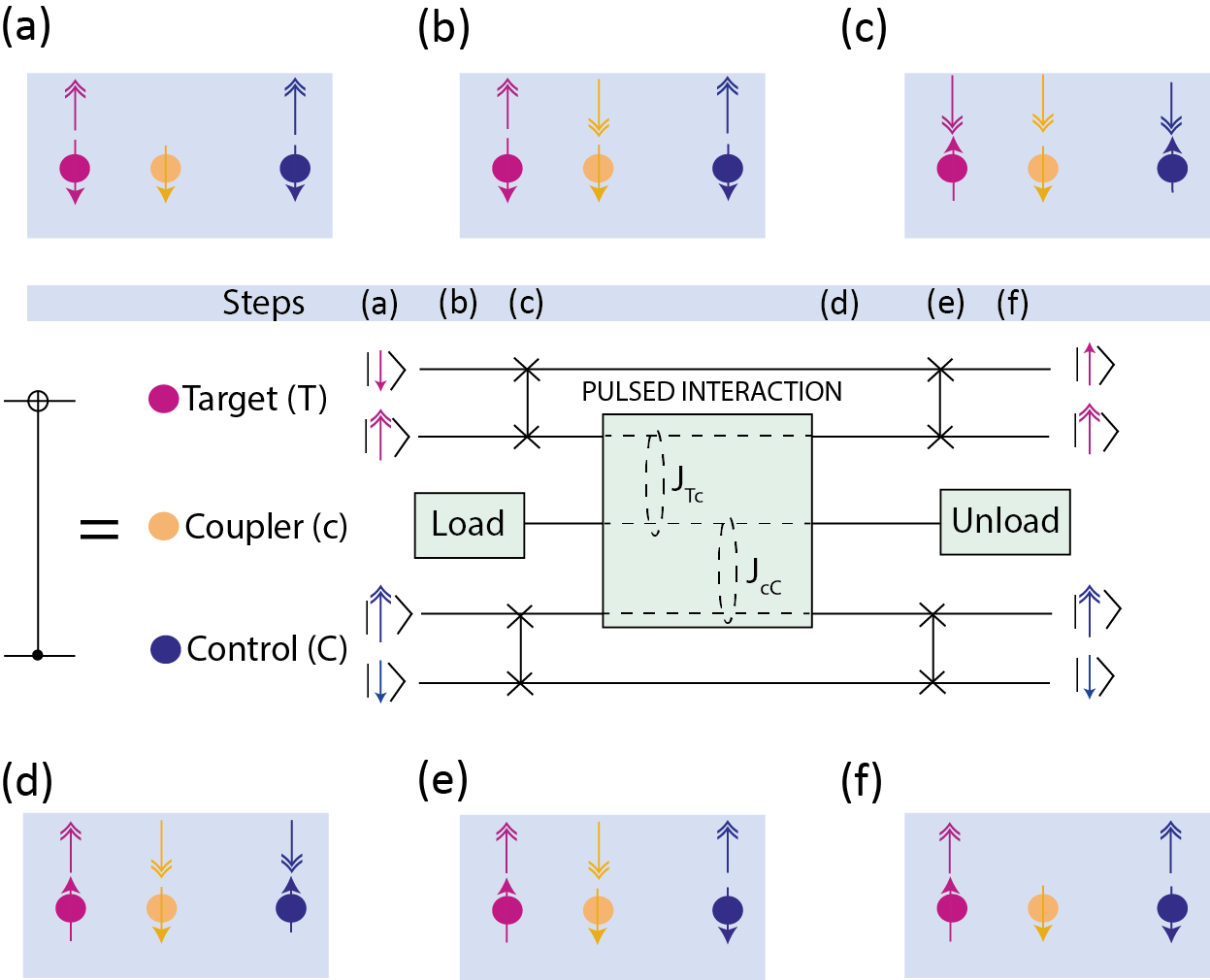}
\caption{\textbf{Coupler-mediated CNOT gate:} A schematic circuit diagram showing conceptual triple-donor CNOT gate construction illustrated for the case $\ket{1}\ket{1} \rightarrow \ket{1}\ket{0}$. In our convention, arrows with single (double) heads label nuclear (electron) spins, and down (up) direction of arrows define $\ket{1}$ ($\ket{0}$). The CNOT gate comprises three phosphorus donor qubit: target (T), control (C), and coupler (c). (a-f) The spin configurations of electron and nuclear spins on three qubits are shown at various stages of the CNOT circuit operation.}
\label{fig:cnot}
\end{center}
\end{figure*}

The design of a robust two-qubit CNOT gate is a fundamental component of any quantum computer architecture. Figure~\ref{fig:cnot} plots the schematic diagram (center circuit) of our design for a two-qubit CNOT gate based on the coupler qubit, digitally controlling the exchange interaction between the control and target data qubits. This mechanism allows the placement of control and target qubits at distances commensurate with the pitch limit of STM control lines and yet achieve MHz to GHz exchange interactions mediated via the coupler qubit. The operation sequence of the proposed CNOT gate is explained in steps (a) to (f) as shown in the diagram. We have indicated both the nuclear and electron qubit spins on each qubit by plotting single and double head arrows, respectively. As shown in (a), we assume that the gate is initialised as both electron spins on control and target qubits in down-spin configuration and the nuclear spins encode the qubit information. In the second step, (b), the coupler qubit is loaded with an electron in down-spin configuration. Next, (c), the nuclear and the electron spins of the target and control qubits are swapped. The CNOT operation is performed between the target and control qubits (d), and then the electron/nuclear spins are swapped again (e) to store the information back in the nuclear spins. Finally, (f) brings the circuit back to the initial condition by unloading the electron from the coupler qubit. This will turn off the interaction between the target and control qubits.     

\begin{figure*}[htbp]
\begin{center}
\includegraphics[width=17cm]{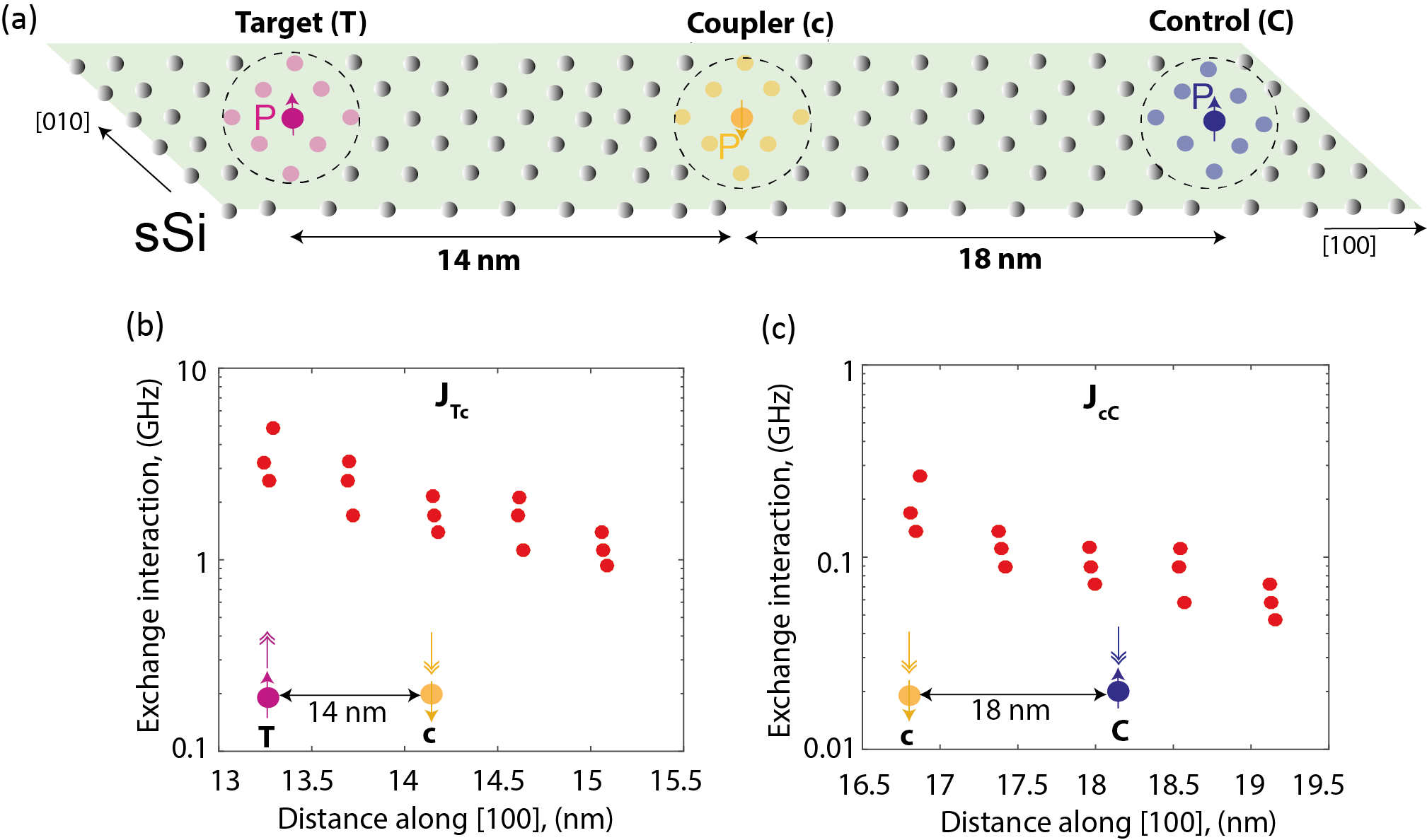}
\caption{\textbf{Exchange distributions for triple donor protocol:} (a) The possible spatial locations are shown within the $\pm a_0$ placement precision for the target (T), coupler (c) and control (C) dopants. Each dopant atom could be placed on one of the possible nine locations, resulting in 81 values for exchange interaction $J_{\rm Tc}$ and $J_{\rm cC}$. However, due to silicon crystal symmetry, only 15 configurations are distinct. (b, c) The distinct values of exchange interactions $J_{\rm Tc}$ and $J_{\rm cC}$ are plotted for 14 nm and 18 nm separations selected between target/coupler and coupler/control, respectively.}
\label{fig:J_dist}
\end{center}
\end{figure*}
   
\subsection{Exchange strength and distribution}
\noindent
The current state-of-the-art scanning tunnelling microscope (STM) based atomic-precision fabrication technology~\cite{Fuechsle_NN_2012} has demonstrated donor placement with $\pm a_0$ accuracy, where $a_0$ is the lattice constant of silicon. However, even such small variations in the donor position may lead to considerably large variations in exchange interaction \cite{Cullis_PRB_1970, Wellard_PRB_2003, Gonzalez_Nanoletters_2014, Hu_PRB_2005, Sarma_SSC_2005, Wellard_JPCM_2004, Kettle_PRB_2006, Kettle_JPCM_2004, Koiller_PRB_2004, Song_APL_2016, Wellard_Hollenberg_PRB_2005, Testolin_PRA_2007, Saraiva_arx_2014, Pica2_PRB_2014, Koiller_Adabdc_2005, Voisin_2020, Usman_CMS_2021}, placing stringent requirement on uniformity assumptions in the design of control schemes for large-scale architectures \cite{Testolin_PRA_2007, Usman_CMS_2021}. In the past, strategies have been developed to mitigate the impact of exchange variations, which include the design of robust composite pulse schemes such as BB1~\cite{Hill_PRL_2007}, exchange characterisation~\cite{Testolin_PRA_2007}, the application of electric fields~\cite{Wang_NPJQ_2016} and the placement of donor atoms along the [110] crystal direction~\cite{Voisin_2020}. In this work, we propose the application of a small strain field (5\%) which allows full control of exchange interaction variation for both in-plane and out-of-plane donor position variations~\cite{Usman_CMS_2021}.  

Fig.~\ref{fig:J_dist} (a) plots a schematic illustration of donor positions for target, coupler and control qubits. Each qubit is indicated by the target donor position and the possible locations under $\pm a_0$ placement imprecision, which is commensurate with the precision placement of donor atoms by STM fabrication techniques. This results in 81 possible configurations between target and coupler, and likewise another 81 possible configurations between coupler and control. We note that due to the symmetry of the silicon crystal lattice, only 15 configurations out of the 81 possibilities are distinct.

The calculation of exchange interaction is performed based on the atomistic tight-binding wave functions of donor electrons in silicon~\cite{Usman_JPCM_2015, Usman_PRB_2015} and the Heiltler-London theory~\cite{Wellard_Hollenberg_PRB_2005}. Fig.~\ref{fig:J_dist} (c) and (d) plots the computed exchange values for the 15 distinct donor configurations between the target and coupler and between the coupler and control, respectively. As an example, the separations between the target and the coupler qubits is selected as 14 nm, and between the coupler and control qubits as 18 nm. These separations allow a pitch of 32 nm which is consistent with the reported STM control-line requirements ($\geq$ 30 nm)~\cite{Hill_science_2015}. We note that the two separations are purposely selected to be slightly different (18 nm and 14 nm), to minimise frequency band overlaps which will allow efficient design of control pulses addressing individual donor pairs.

Figure~\ref{fig:J_dist}(c) and (d) show a relatively small variation in exchange interaction (about a factor of 5 or less), when compared to roughly three orders of magnitude variation reported for similar donor position uncertainties in unstrained silicon substrate~\cite{Voisin_2020}. This considerably suppressed variation in exchange strength has important implication for the fidelity of CNOT gate which sharply decreases when the exchange distribution is large~\cite{Testolin_PRA_2007}. Furthermore, full exchange control can be achieved in strained silicon system by an external in-plane electric field which can provide a tuning of factor or ten or more for donor separations above 14 nm~\cite{Usman_CMS_2021}.  

\begin{figure*}[htbp]
\begin{center}
\includegraphics[width=14cm]{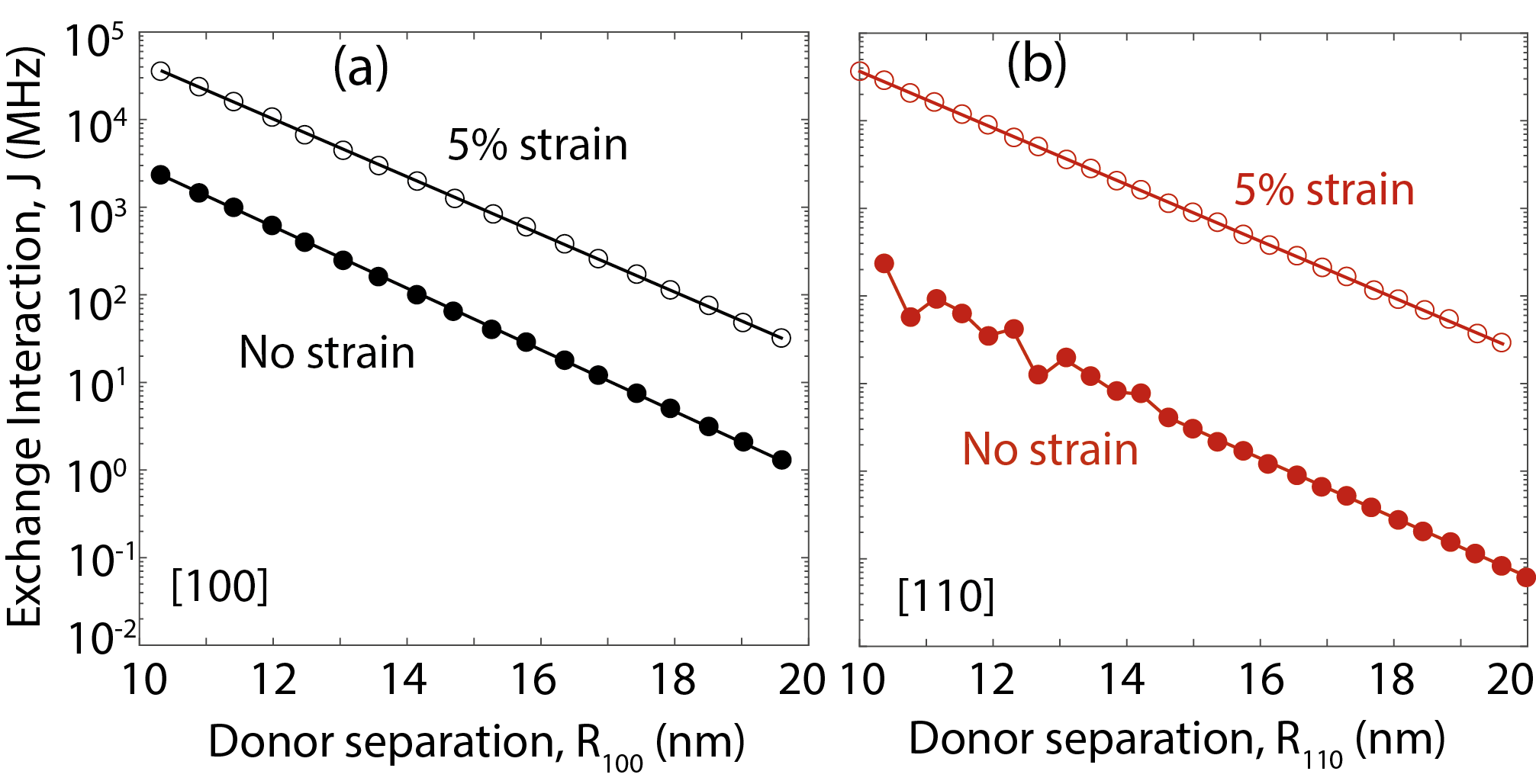}
\caption{\textbf{Exchange enhancement:} (a, b) Exchange interactions ($J$) between two P atoms separated along the [100] and [110] directions are plotted for both unstrained (diamond symbols) and 5\% strained (square symbols) silicon substrates. The $J$ values are presented in the exchange term of the effective spin Hamiltonian ($J \vec{\sigma^e_1} \cdot \vec{\sigma^e_2}$), in which case $J$ = $\frac{E_T - E_S}{4}$, where $E_T - E_S$ is the singlet-triplet splitting. The conversion of energy to frequency is based on 1 meV $\sim$ 242 MHz. }
\label{fig:exchange_enhancement}
\end{center}
\end{figure*}

The application of strain offers another direct benefit in terms of CNOT gate operation times as the interaction time is inversely proportional to exchange strength. Figure~\ref{fig:exchange_enhancement} plots exchange strength for various donor separations along the [100] and [110] directions for both unstrained and strained silicon environments. From these plots, a two-fold impact of strain is evident. First, the application of strain significantly boosts the strength of exchange interaction, as also reported in the literature \cite{Wellard_Hollenberg_PRB_2005, Koiller_PRB_2002, Koiller_PRB_2004, Sarma_SSC_2005, Kettle_PRB_2006}. For example, our calculations show that donors placed at 20 nm separation in strained silicon will have roughly the same exchange interactions as the donor pairs which are 12-14 nm separations in the unstrained silicon. This implies that donors can be placed much larger distances in strained system without sacrificing exchange interaction or CNOT interaction times, which is important to meet the pitch requirements of a large-scale architecture. From our calculations, we estimate O($\upmu$sec) interaction times for donor separations of upto 25 nm in strained silicon case, which is drastically faster when compared to O($m$sec) interaction times for unstrained silicon substrates. Secondly, the exchange interaction in strained environment is highly uniform, i.e., nearly same strength along the [100] and [110] directions. The uniformity of exchange strength with respect to donor placement orientation ([100] and [110]) will be useful in the design of a planar 2D surface-code architecture such as proposed in this work (Figure~\ref{fig:architecture}).      

\subsection{GRAPE Pulse Engineering}
\noindent
The configurations of donor separations as shown in Figure~\ref{fig:J_dist} lead to a distribution of corresponding interaction strengths, $J_{Tc}$ and $J_{cC}$. Typically, at the selected spacing of 14-18 $\mathrm{nm}$ these coupling strengths are larger than the hyperfine interaction, $A$,  and so do not fall into the regime described in Figure~\ref{fig:cnot}. Conceptually, the same operations are being applied, however since all three electrons are strongly interacting, the control pulses do not lend themselves to such a simple description. In order to quantitatively determine control pulses required, we calculated pulses for the electron to electron CNOT gate from control to target electrons using numerically optimized GRAPE sequences.

\begin{figure*}[htbp]
\begin{center}
\includegraphics[width=18cm]{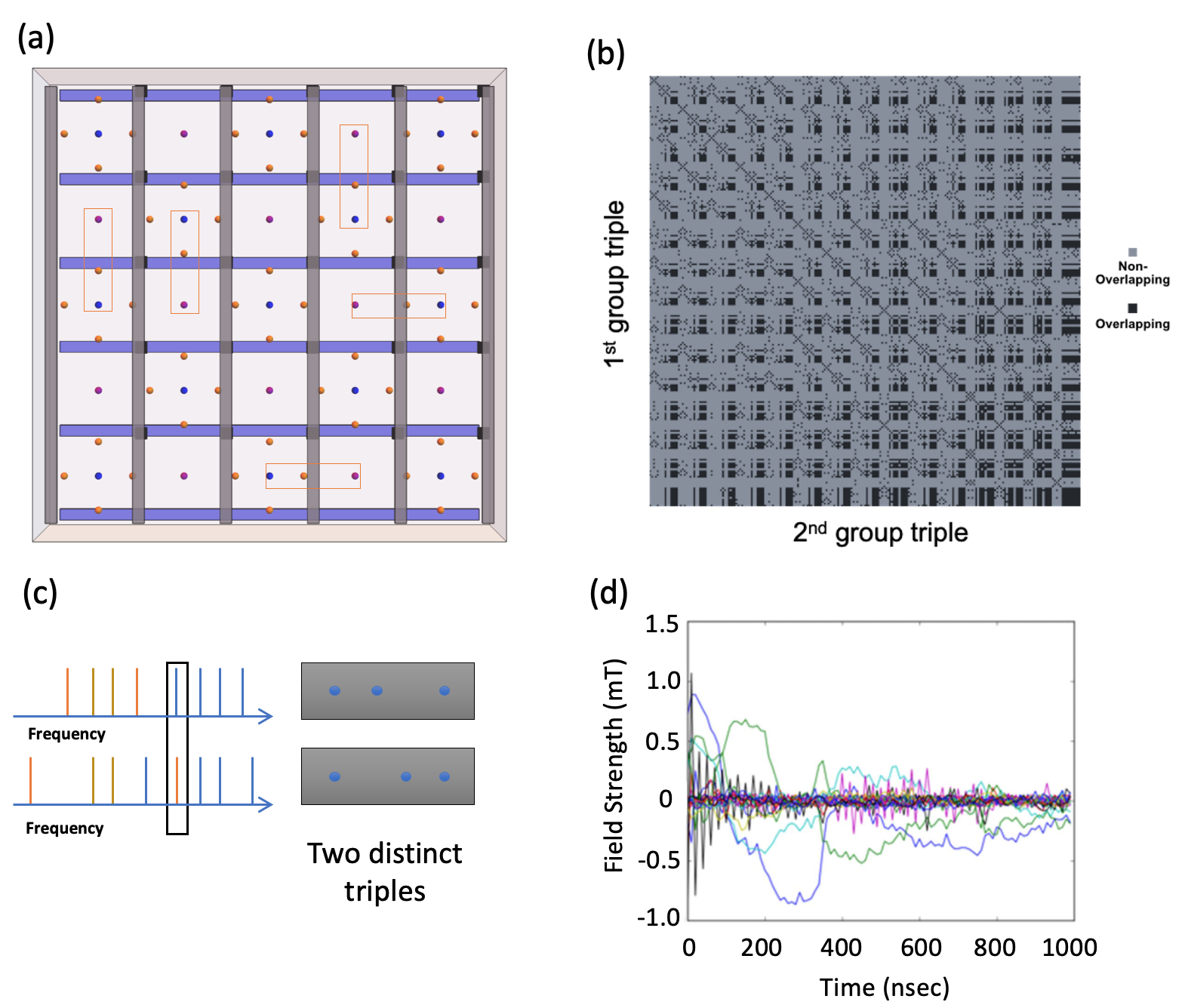}
\caption{\textbf{Engineered Pulse Control:} Schematic showing the strategy for developing control pulses for a large array of donors. (a) The placement of donors gives rise to different transition frequencies (b) Several of these frequencies will overlap between distinct donor triples. (c) From these donor triples, we identify sets of potential candidates triples for concurrent pulses - spatially separated and either non-overlapping transitions in frequency space, or with frequencies amenable to a broadband pulse (d) Optimal pulses are found numerically using GRAPE which concurrently applies a CNOT to all donor triples in that set. Difference colors indicate optimised pulse sequences for different frequency combinations.}
\label{fig:overlap}
\end{center}
\end{figure*}

Since a wide range of exchange interaction strengths would be present in our architecture, our strategy for implementing these pulses started from a simple electron spin Hamiltonian (in 
the absence of an $AC$ control pulse applied):

\begin{eqnarray}
    H_{\rm en} &=& g \mu_B B (Z_T + Z_C + Z_{c}) + g_n \mu_n B (Z_{nT} + Z_{nC} + Z_{nc}) \nonumber \\
    && + A_T \sigma_T \cdot \sigma_{nT}  + A_C \sigma_C \cdot \sigma_{nC} + A_{c} \sigma_{c} \cdot \sigma_{nc} \nonumber \\
    && + J_{Tc} \sigma_T \cdot \sigma_c + J_{cC} \sigma_c \cdot \sigma_{C}
\end{eqnarray}
\noindent
where $T$, $C$, and $c$ subscripts refer to the electron spins corresponding to target, coupler and control qubits respectively, and the corresponding $nT$, $nC$, and $nc$ refer to the nuclear spins. Here, and throughout the paper, $X$, $Y$ and $Z$ are the Pauli spin operators, and $\sigma \cdot \sigma$ the exchange interaction between spins. Using the approximation that nuclear spins remain static during this evolution, the electron spin Hamiltonian can be reduced to the more tractable,

\begin{eqnarray}
    H_{\rm e} &=& (g \mu_B B +A_T) Z_T + (g \mu_B B - A_C) Z_C + (g \mu_B B +A_{c}) Z_{c}  \nonumber \\
    && + J_{Tc} \sigma_T \cdot \sigma_c + J_{cC} \sigma_c \cdot \sigma_{C} \label{eqn:fullH}
\end{eqnarray}

We wish to control the electron spins with a transverse $AC$ field,

\begin{eqnarray}
    H_{\rm AC} &=& g \mu_B B_{AC} \cos{\omega_r t} \left(X_T + X_c + X_C \right) \nonumber 
    \\ && + g \mu_B B_{AC} \sin {\omega_r t} \left( Y_T + Y_c + Y_C \right) \label{eqn:Hac} 
\end{eqnarray}
where typically $\omega$ is chosen to be resonant with a transition between two of the eigenstates of $H$ given in Eqn. (\ref{eqn:fullH}).

Not every transition between every pair of eigenstates is allowed. As an illustrative example, if $J_{Tc} \gg A$ and $J_{Tc} \gg J_{cC}$ then a transverse field of the form of Eqn. (\ref{eqn:Hac}) would not excite transitions between the singlet and triplet eigenstates due to symmetry considerations. Note, however, that over a long time period, even though an individual transition might not be able to be individually addressed, the symmetry can be broken because the central spin interacts with both neighbours. Such disallowed transitions can be identified numerically by considering the off-diagonal elements of $H_{AC}$ given in Eqn. (\ref{eqn:Hac}) written in the eigenbasis of $H_{e}$ given in Eqn. (\ref{eqn:fullH}). In addition, two transitions can lie close in frequency, and not able to be individually addressed in experiment. These two considerations given rise to a viable set of control frequencies, $\omega$ which significantly excite transition between eigenstates of $H_{e}$ and can be effectively addressed in experiment.

We performed GRAPE numerical optimization to determine gate pulse sequences for the CNOT gate between electron spins. To do this, we considered each of the different resonant frequencies which excite transitions between eigenstates of the system as different control parameters. At each time-step, it was possible to vary the strength of the $AC$ field applied, as well as the phase of the applied microwave field. Using gradient ascent, we optimized the trace fidelity,

\begin{equation}
    F(U) = \mathrm{Tr} \left[ U_{C} U_{G} \right]
\end{equation}
where $U_C$ is the perfect CNOT gate applied between electronic spin states 1 and 3 and leaving the second electronic spin unchanged. $U_G$ is the obtained evolution obtained from a given GRAPE pulse sequence.

We repeated GRAPE for each of the 225 different pairs of strengths of exchange interactions $J_{Tc}$ and $J_{cC}$, obtaining a numerically optimized CNOT pulse sequence in each case. Almost all pulse sequences resulted in a high fidelity CNOT gate, accurate to $0.1\%$ accuracy. Only six CNOT gates had lower fidelities. We note that there are 225 different triples of qubits. To operate on each of these triples independently would require 225 different pulse schemes - such as those calculated by GRAPE. However, many of these pulses can, in principle, be applied in parallel. This can be applied in parallel if (i) pulses have disjoint frequencies, which do not overlap, (ii) broadband pulses can be applied to implement the gate on triples with near-lying frequencies.

Pulses with disjoint frequencies can be operated in parallel, since an out of resonance field will not excite transitions in off-resonant spins. The larger the number of triples with non-overlapping frequencies, the more operations that can be applied in parallel because they have disjoint frequencies. A rough estimate of the number of triples (CNOT gates) that can be made is as follows: If any two triples have a probability of 30\% (40\%) of having a transition with an overlapping frequencies, then approximately 12 (9) of the 225 CNOT gates can be chosen to operate in parallel. Further tuning of exchange interactions can be performed by the application of external electric fields~\cite{Usman_CMS_2021}, which could allow more frequencies to be operated in parallel.

\section{Summary}
\noindent
We have introduced a new concept for the incorporation of fast exchange interaction in surface-code architecture scheme for donor spin qubits in silicon. The proposal is underpinned by the design of a CNOT gate in which the coupling between target and control data qubits in mediated by an additional coupler qubit which can selectively turn on/off exchange interaction between data qubits. The introduction of coupler qubit allows data qubits to be placed at large separations ($\geq$ 30 nm) commensurate with the requirements of a large-scale architecture. We also discuss the application of a small strain field (~5\%) which provides important benefits such as significant enhancement in exchange strength leading to O($\upmu$sec) interaction times, suppressed exchange variation arising from the donor placement inaccuracy and uniformity in exchange interactions along the [100] and [110] crystal directions. We consider a both global control as well as targeted GRAPE control based on mapping frequency distributions arising from exchange variations. The work here is a step on the path to the design and implementation of a large-scale error-corrected quantum computer architecture based on atomic spin qubits in silicon.

\noindent
\\ \\
\textbf{Acknowledgements:} This work was supported by the Australian Research Council (ARC) funded Center for Quantum Computation and Communication Technology (CE170100012). Computational resources were provided by the National Computing Infrastructure (NCI) and Pawsey Supercomputing Center through National Computational Merit Allocation Scheme (NCMAS). This research was undertaken using the LIEF HPC-GPGPU Facility hosted at the University of Melbourne. This Facility was established with the assistance of LIEF Grant LE170100200.
\\ \\
\noindent
\textbf{Author contributions:} All authors contributed in the development of the concept, planning, data analysis and writing of the manuscript. 
\\ \\
\noindent
\textbf{Conflict of Interest:} The authors declare no competing financial or non-financial interests. A patent application has been filed based on aspects of the architecture design.
\\ \\
\noindent
\textbf{Data availability:}
The data that support the findings of this study are available within the article. Further information can be provided upon reasonable request. 

\def\bibsection{\subsection*{\refname}}

\bibliographystyle{apsrev4-1}

%

\end{document}